\documentclass[journal=jpclcd,manuscript=letter]{achemso}
\usepackage{amsmath,amssymb}
\usepackage{float}
\usepackage[version=3]{mhchem} 
\usepackage{graphicx}
\usepackage{adjustbox}
\usepackage{amsmath,amssymb}
\usepackage{mathrsfs}
\usepackage{color}
\usepackage{multirow}
\usepackage{xcolor}
\usepackage{array}
\usepackage{afterpage}
\usepackage{upgreek}
\usepackage{enumerate}
\usepackage[normal]{subfigure}
\usepackage[figurename={Figure}]{caption}
\usepackage{float}
\usepackage{gensymb} 

\usepackage{titlesec}
\titlelabel{\thetitle. }

\SectionNumbersOn

\author{Preeti Bhumla}
\email{Preeti.Bhumla@physics.iitd.ac.in [PB]}
\author{Deepika Gill}
\author{Sajjan Sheoran}
\author{Saswata Bhattacharya}
\email{saswata@physics.iitd.ac.in [SB]}
\phone{+91-11-2659 1359}
\fax{+91-11-2658 2037}
\affiliation[Indian Institute of Technology Delhi]
{Department of Physics, Indian Institute of Technology Delhi, New Delhi, India}
\title[An \textsf{achemso} demo]
{Origin of Rashba Spin-Splitting and Strain Tunability in Ferroelectric Bulk CsPbF$_3$}
\begin{document}
\begin{center}
{\Large \bf Supplemental Material}\\ 
\end{center}
\begin{enumerate}[\bf I.]
	
\item Optimized structures of \textit{Pm}$\bar{3}$\textit{m} and \textit{R}${3}$\textit{c} phases.	
\item Band structure and projected density of states (pDOS) of cubic \textit{Pm}$\bar{3}$\textit{m} phase.
\item Phonon band structures of \textit{Pm}$\bar{3}$\textit{m} and \textit{R}${3}$\textit{c} phases.
\item  \textit{k}-grid convergence in \textit{Pm}$\bar{3}$\textit{m} and \textit{R}${3}$\textit{c} phases.  	
\item Imaginary part (Im($\varepsilon$)) of the dielectric function for cubic \textit{Pm}$\bar{3}$\textit{m}  phase.
\item Comparison of PBE+SOC, HSE06+SOC and G$_0$W$_0$@HSE06+SOC band structures of cubic \textit{Pm}$\bar{3}$\textit{m} phase.
\item Higher order term contribution in SOC Hamiltonian of \textit{R}${3}$\textit{c} phase.
\item RD parameters as a function of strain in \textit{R}${3}$\textit{c} phase.
\item Band gap as a function of strain in \textit{R}${3}$\textit{c} phase.

\end{enumerate}
\vspace*{12pt}
\clearpage

\section{Optimized structures of \textit{Pm}$\bar{3}$\textit{m} and \textit{R}${3}$\textit{c} phases}
\begin{table}[htbp]
	\caption {Structure of optimized  \textit{Pm}$\bar{3}$\textit{m} CsPbF$_3$, a=4.80652 \AA}
	\begin{center}
		\begin{adjustbox}{width=0.32\textwidth} 
			\setlength\extrarowheight{+1pt}
			\begin{tabular}[c]{|c|c|c|c|c|}  \hline		
				\textbf{Atom} & \textbf{x} & \textbf{y} & \textbf{z}  \\ \hline	
				Cs (1a)      &  0  & 0 & 0   \\ \hline
				Pb (1b)     &  0.5  &  0.5  & 0.5  \\ \hline
				F (3c)     &  0  & 0.5 & 0.5     \\ \hline

			\end{tabular}
		\end{adjustbox}
		\label{T3}
	\end{center}
\end{table}
\begin{table}[htbp]
	\caption {Structure of optimized \textit{R}${3}$\textit{c} CsPbF$_3$, a=6.80864 \AA, c=16.30620 \AA. Hexagonal settings are used.}
	\begin{center}
		\begin{adjustbox}{width=0.4\textwidth} 
			\setlength\extrarowheight{+1pt}
			\begin{tabular}[c]{|c|c|c|c|c|}  \hline		
				\textbf{Atom} & \textbf{x} & \textbf{y} & \textbf{z}  \\ \hline	
				Cs (6a)      &  0  & 0 & 0   \\ \hline
				Pb (6b)     &  0 &  0.5  & 0.5  \\ \hline
				F (18e)     &  0.44072  & 0 & 0.25     \\ \hline

			\end{tabular}
		\end{adjustbox}
		\label{T3}
	\end{center}
\end{table}
\newpage

\section{Band structure and projected density of states (pDOS)  of cubic \textit{Pm}$\bar{3}$\textit{m}}
\begin{figure*}[h!]
	\includegraphics[width=1.0\columnwidth,clip]{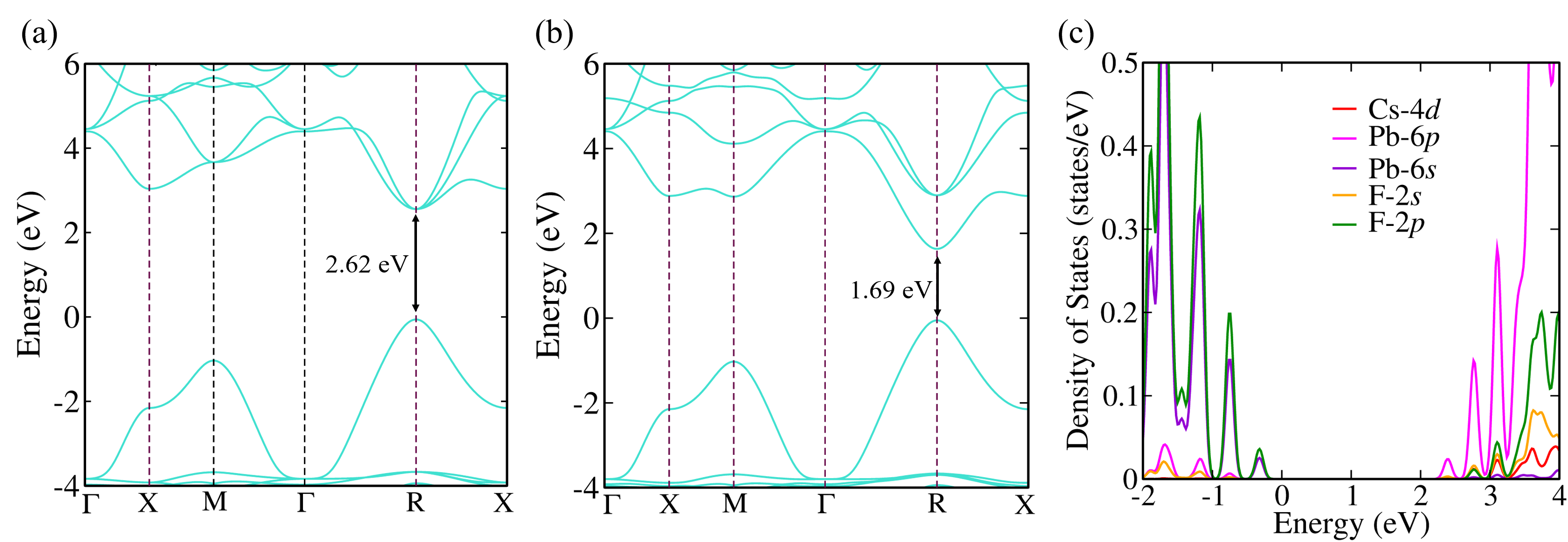}
	\caption{Band structure of  cubic \textit{Pm}$\bar{3}$\textit{m} phase of CsPbF$_3$, calculated using (a) PBE (b) PBE+SOC. (c) Projected density of states (pDOS), calculated using HSE06+SOC.}
	\label{figS1}
\end{figure*}
\newpage

\section{Phonon band structures of \textit{Pm}$\bar{3}$\textit{m} and \textit{R}${3}$\textit{c} phases}
\begin{figure}
\includegraphics[width=11cm]{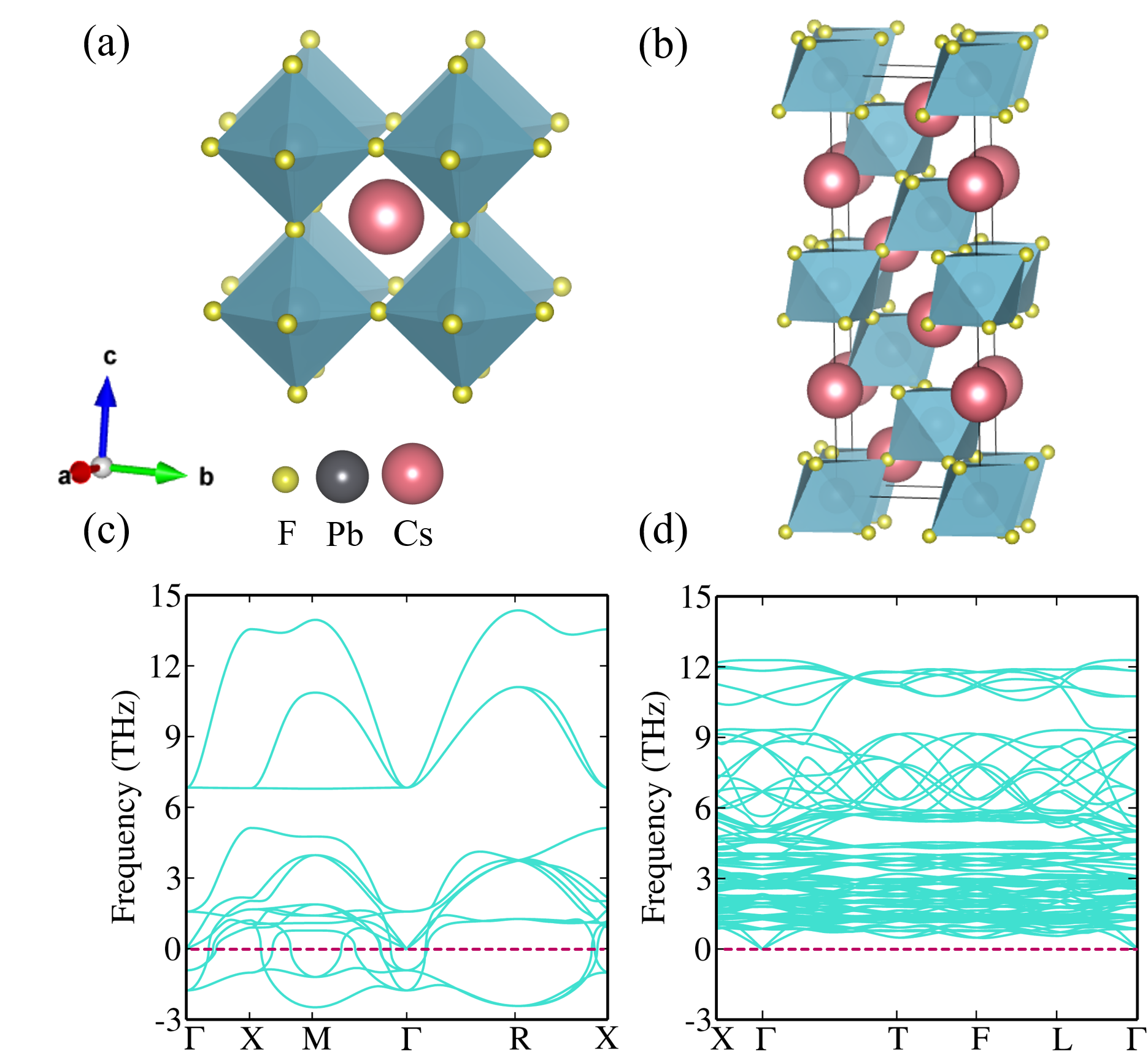}
\caption{Crystal structure of CsPbF$_3$ (a) cubic \textit{Pm}$\bar{3}$\textit{m} phase and (b) rhombohedral \textit{R}3\textit{c} phase. Cs, Pb and F atoms are indicated by red, grey and yellow colors, respectively. Phonon band structure of CsPbF$_3$ (c) cubic \textit{Pm}$\bar{3}$\textit{m} phase and (d) rhombohedral \textit{R}3\textit{c} phase.}
\label{1}
\end{figure}
The optimized crystal structures of \textit{Pm}$\bar{3}$\textit{m} and \textit{R}3\textit{c} phases are shown in Fig. \ref{1}(a) and \ref{1}(b). To analyze the dynamical stability, we have plotted the phonon band structures for optimized \textit{Pm}$\bar{3}$\textit{m} and \textit{R}3\textit{c} phases, respectively. Fig. \ref{1}(c) and \ref{1}(d) show the phonon band structures calculated along the high-symmetry points of the Brillouin zone. Notably, the negative frequencies can be seen in phonon band structure of cubic phase (see Fig. \ref{1}(c)) leading to the instability of this phase. The absence of negative frequencies in Fig. \ref{1}(d) confirms the dynamical stability of \textit{R}3\textit{c} phase.
\newpage
\section{\textit{k}-grid convergence in \textit{Pm}$\bar{3}$\textit{m} and \textit{R}${3}$\textit{c} phases}
\begin{figure}[h!]
	\includegraphics[width=0.8\columnwidth,clip]{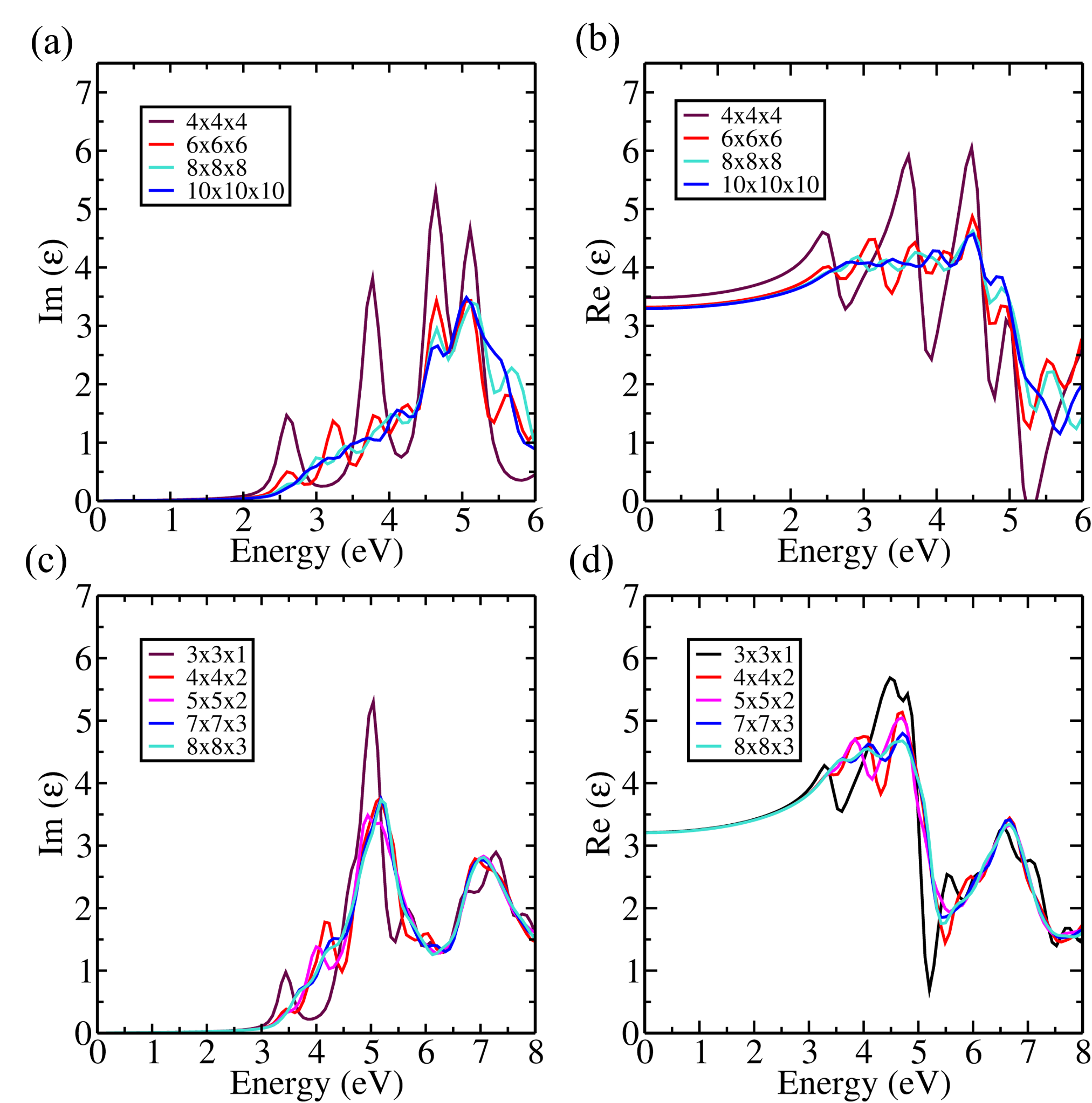}
	\caption{Real (Re($\varepsilon$)) and imaginary (Im($\varepsilon$)) part of dielectric function for \textit{Pm}$\bar{3}$\textit{m} and \textit{R}${3}$\textit{c} phases using PBE $\epsilon_{xc}$ functional.}
	\label{figS2}
\end{figure}
Fig. \ref{figS2} (a) and (b) show the variation of imaginary (Im($\varepsilon$)) and real part (Re($\varepsilon$)) of dielectric function for cubic \textit{Pm}$\bar{3}$\textit{m} phase while Fig. \ref{figS2} (c) and (d) show the imaginary (Im($\varepsilon$)) and real part (Re($\varepsilon$)) of dielectric function for \textit{R}${3}$\textit{c} phase. On increasing \textit{k}-grid, no shift is observed in first peak position of Im($\varepsilon$) part of dielectric function. Hence, 6$\times$6$\times$6 and 4$\times$4$\times$2 \textit{k}-grids are sufficient to compute quasi particle (G$_0$W$_0$) band gap in \textit{Pm}$\bar{3}$\textit{m} and \textit{R}${3}$\textit{c} phases, respectively. 

\section{Imaginary part (Im($\varepsilon$)) of the dielectric function for cubic \textit{Pm}$\bar{3}$\textit{m} phase}
\begin{figure}[h!]
	\includegraphics[width=0.45\columnwidth,clip]{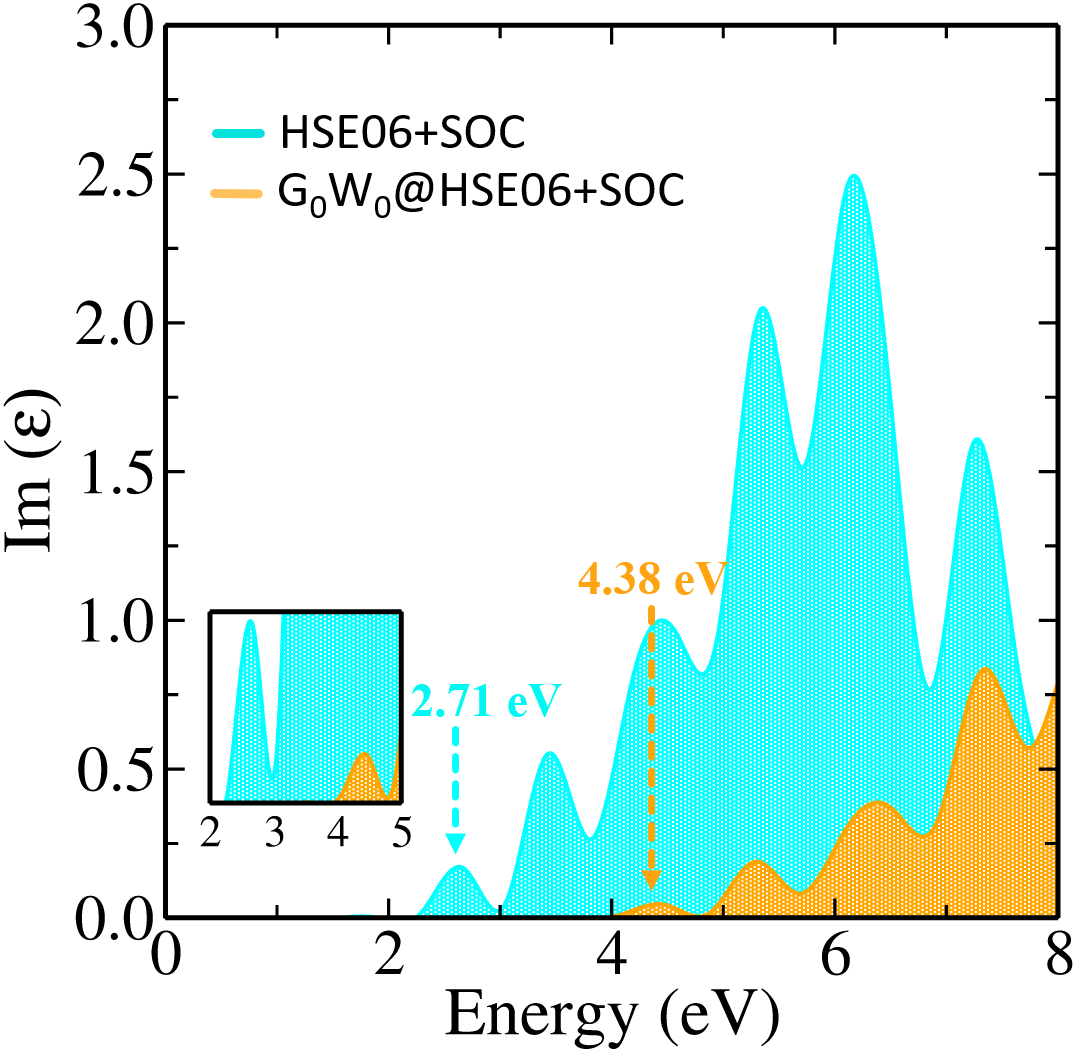}
	\caption{Imaginary part (Im ($\varepsilon$)) of the dielectric function for \textit{Pm}$\bar{3}$\textit{m}  phase of CsPbF$_3$ calculated using HSE06+SOC and G$_0$W$_0$@HSE06+SOC $\epsilon_{xc}$ functional, respectively.}
	\label{figS1}
\end{figure}
\newpage

\section{Comparison of PBE+SOC, HSE06+SOC and G$_0$W$_0$@HSE+SOC band structures of cubic \textit{Pm}$\bar{3}$\textit{m} phase}
\begin{figure}[h!]
	\includegraphics[width=1\columnwidth,clip]{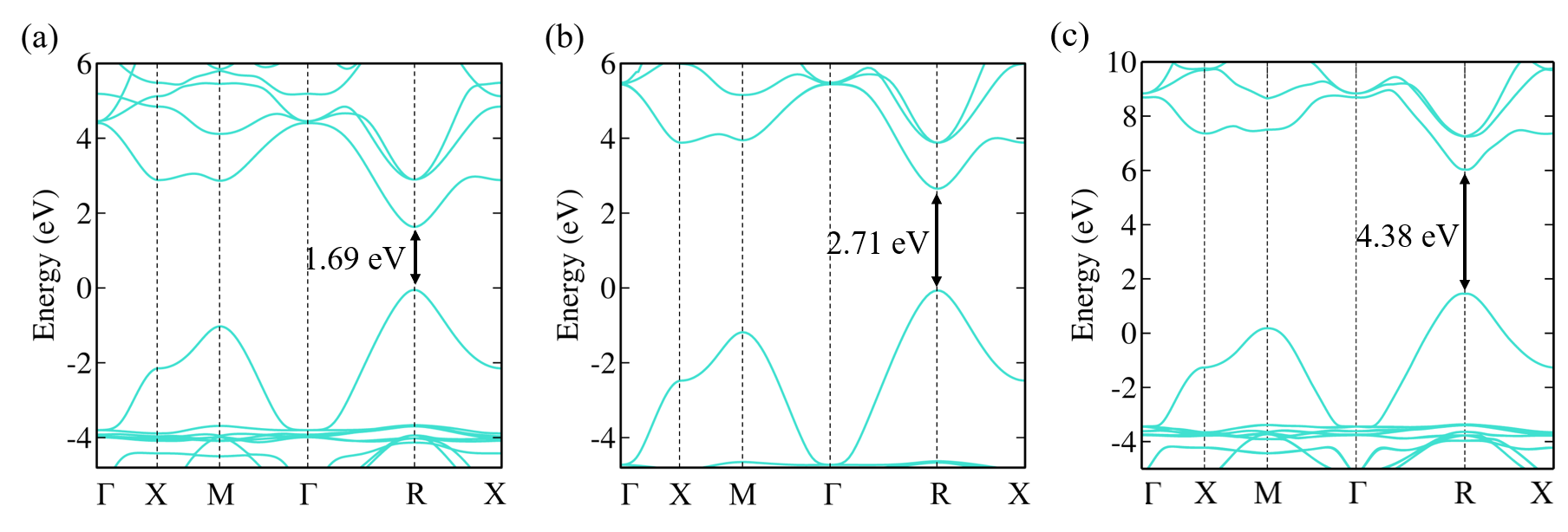}
	\caption{Band structure of  cubic \textit{Pm}$\bar{3}$\textit{m} phase of CsPbF$_3$, calculated using (a) PBE+SOC (b) HSE06+SOC (c) G$_0$W$_0$@HSE06+SOC, respectively.}
	\label{figS1}
\end{figure}
\newpage
\section{Higher order term contribution in SOC Hamiltonian of \textit{R}${3}$\textit{c} phase}
The model Hamiltonian of $R3c$ structure possessing $C_{3v}$ point group symmetry near $\Gamma$ point in the presence of SOC can be written as \cite{PhysRevB.101.014109, PhysRevB.95.245141, PhysRevB.97.205138, sheoran2021rashba}:
\begin{equation}
H_\Gamma(\textbf{\textit{k}})=H_o(\textbf{\textit{k}})+ H_{SO}
\label{Eq_1}
\end{equation}
where, 
\begin{equation}
H_{SO}=\alpha \sigma_y k_x + \beta \sigma_x k_y 
+ \gamma \sigma_z[(k_x^3+k_y^3)-3(k_xk_y^2+k_yk_x^2)]
\end{equation}
and $H_o(\textbf{\textbf{\textit{k}}})$ is free particle Hamiltonian. $\alpha$, $\beta$ are the coefficients  of linear terms and $\gamma$ is the coefficient of cubic term in SOC Hamiltonian. The energy eigenvalues of Hamiltonian are given as:\\
\begin{equation}
E(\textbf{\textit{k}})^{\pm}= \frac{\hslash^2 k_x^2}{2m_x}+\frac{\hslash^2 k_y^2}{2m_y} \pm E_{SO}  
\end{equation}
where, $m_x$ and $m_y$ represent the effective masses in $x$ and $y$ directions, respectively. $E_{SO}$ is the energy eigenvalue of SOC Hamiltonian given by $E_{SO}(\textbf{\textit{k}})=\sqrt{\alpha^2 k_x^2+\beta^2 k_y^2+ \gamma^2 f^2(k_x,k_y)}$, where $f(k_x,k_y)=(k_x^3+k_y^3)-3(k_xk_y^2+k_yk_x^2)$. The normalized spinor wavefunctions corresponding to these energy eigenvalues are given by
\begin{equation}
\Psi_{\textbf{\textit{k}}}^{\pm} = \frac{e^{i\textbf{\textit{k.r}}}}{\sqrt{2\pi(\rho^2_{\pm}+1 })}\begin{pmatrix}
\frac{i\alpha k_x-\beta k_y}{\gamma f(k_x,k_y)\mp E_{SO}} \\
1
\end{pmatrix}
\end{equation}
where $\rho^2_{\pm}=\frac{\alpha^2 k_x^2+\beta^2 k_y^2}{(\gamma f(k_x,k_y)\mp E_{SO})^2}$. The expectation values of spin operators are given by 
\begin{equation}
\{s_x,s_y,s_z\}^{\pm}=\pm \frac{1}{E_{so}}\{\beta k_y,\alpha k_x,\gamma f(k_x,k_y) \}
\end{equation}
The in-plane spin components are given by $\alpha$ and $\beta$, whereas out-of-plane spin component can be explained using $\gamma$. The out-of-plane spin component is negligible if $\gamma$ is much smaller than $\alpha$ and $\beta$. Since we have not noticed significant out-of-plane spin component in the spin texture, therefore, we had neglected the higher order terms in the Hamiltonian.To confirm the same, we have calculated the value of $\gamma$ by fitting the model Hamiltonian. The value of $\gamma$ came out to be 0.012 eV$\textrm{\AA}$, which is negligible in comparison to $\alpha$ and $\beta$. 
\section{RD parameters as a function of strain in \textit{R}${3}$\textit{c} phase}
\begin{table}[htbp]
	\caption {RD parameters as a function of strain for CBm in \textit{R}${3}$\textit{c} phase.}
	\begin{center}
		\begin{adjustbox}{width=0.85\textwidth} 
			\setlength\extrarowheight{+4pt}
			\begin{tabular}[c]{|c|c|c|c|c|c|} \hline		
				\textbf{Strain (\%)} & \textbf{$\delta E$ (meV)} & \bf{$\delta$\textit{k}$_{\Gamma-\textrm{M}}$ (\AA$^{-1}$)} & \bf{$\delta$\textit{k}$_{\Gamma-\textrm{K}}$ (\AA$^{-1}$)} & \bf{$\alpha_R$ (eV\AA)} & \bf{$\alpha_D$} \bf{(eV\AA)} \\ \hline
				-5      &  109.2  & 0.126 & 0.103& 1.48     &        0.25       \\ \hline
				-4      &  96.9  &  0.118  & 0.096 & 1.41     &        0.23        \\ \hline
				-3       &  83.5  & 0.108 & 0.087& 1.35     &        0.21      \\ \hline
				-2       &  70.2  &  0.098  & 0.079 & 1.25     &        0.19        \\ \hline
				-1      &  56.8 & 0.086 & 0.070& 1.13     &        0.18       \\ \hline
				0      &  43.5  &  0.075  & 0.061 & 1.00     &        0.16       \\ \hline
				1       &  31.0  & 0.062 & 0.050 & 0.87     &        0.13       \\ \hline
				2       &  21.6  &  0.051  & 0.041 & 0.74     &        0.11        \\ \hline
				3       &  11.9  & 0.037 & 0.030 & 0.55    &        0.09      \\ \hline
				4       &  8.7  &  0.032  & 0.025 & 0.49     &        0.05        \\ \hline
				5       &  3.9  & 0.022 & 0.017& 0.32     &        0.02       \\ \hline
				
			\end{tabular}
		\end{adjustbox}
		\label{T3}
	\end{center}
\end{table}

\newpage
\section{Band gap as a function of strain in \textit{R}${3}$\textit{c} phase.}
\begin{figure}[h!]
	\includegraphics[width=0.7\columnwidth,clip]{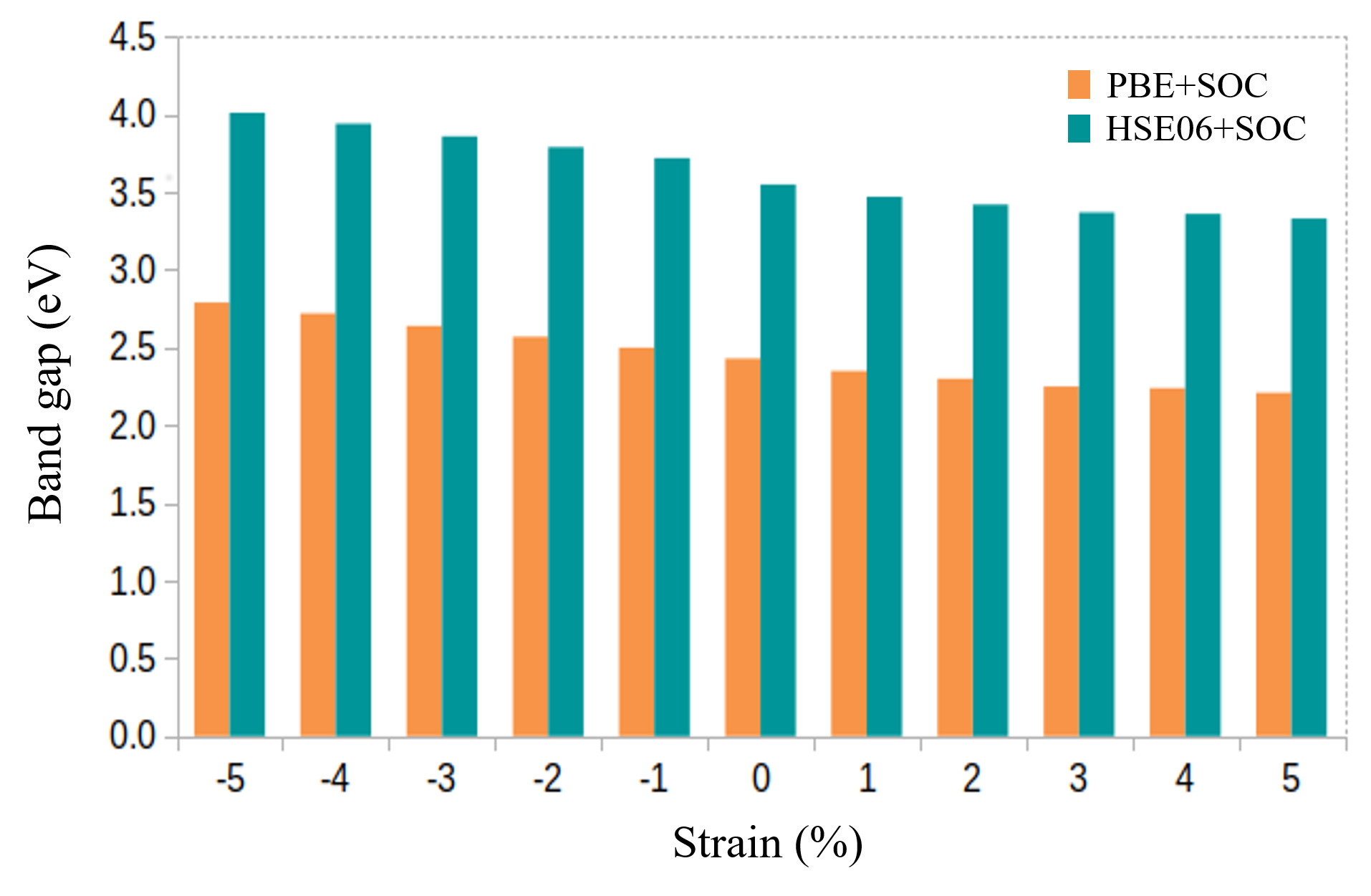}
	\caption{Band gap as a function of uniaxial strain ($\pm$5\%), calculated using PBE+SOC and HSE06+SOC.}
	\label{figS3}
\end{figure}
\newpage

\bibliography{SI.bib}
\end{document}



\begin{abstract}
	
Spin-orbit coupling (SOC) in conjunction with broken inversion symmetry acts as a key ingredient for several intriguing quantum phenomena viz. persistent spin textures, topological surface states and Rashba-Dresselhaus (RD) effects. The coexistence of spontaneous polarization and the RD effect in ferroelectric (FE) materials enables the electrical control of spin degree of freedom. In light of this, we explore here the FE lead halide perovskite viz. CsPbF$_3$ as a potential candidate in the field of spintronics by employing state-of-the-art first-principles based methodologies viz. density functional theory (DFT) with semi-local and hybrid functional (HSE06) combined with SOC and many-body perturbation theory (G$_0$W$_0$). For a deeper understanding of the observed spin-splitting, we have analyzed the spin textures within the combined framework of DFT and $\textbf{\textit{k.p}}$ model Hamiltonian. The latter confirms that there is no out of plane spin component indicating that the Rashba splitting dominates over Dresselhaus splitting in this system. Owing to the presence of Pb-6$p$ orbital in conduction band, the large value of Rashba coefficient ($\alpha_R$) at conduction band minimum (CBm) is noticed in comparison to that of at the valence band maximum (VBM). We also observe that strength of Rashba spin-splitting can be substancially tuned on application of uniaxial strain ($\pm5\%$). More interestingly, we notice reversible spin textures by switching the FE polarization in CsPbF$_3$ perovskite. These findings will further pave the path for perovskite-based spintronics devices.
  \begin{tocentry}
  \begin{figure}[H]%
  	\includegraphics[width=1.0\columnwidth,clip]{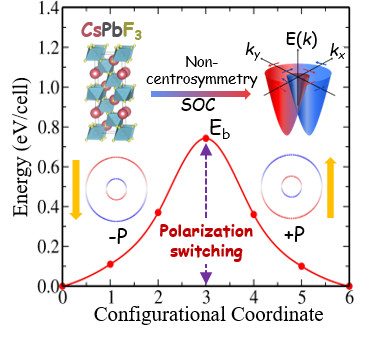}
  \end{figure}	
  \end{tocentry}
\end{abstract}
 Lead halide perovskites 
 have evolved as an excellent choice in optoelectronics owing to their exotic properties viz. suitable optical band gap, high absorption coefficient, low trap density and reasonable manufacturing cost \cite{snaith2013perovskites,green2014emergence,gratzel2014light,kojima2009organometal,lee2012efficient,xing2013long,manser2016intriguing,swarnkar2016bd,stranks2015metal}. These alluring materials exhibit great applications as an absorber for high power conversion efficiency solar cells \cite{nam2018methodologies,tombe2017optical,burschka2013sequential}. Since last decade, consistent research efforts have led to the revolutionary growth of power conversion efficiency to over 25\% \cite{nrel,sahli2018fully}. Additionally, the large spin-orbit coupling (SOC) tied to the presence of heavy element Pb plays a decisive role in determining the electronic properties of lead halide perovskites \cite{even2012electronic,even2013importance}. Over the years, the influence of SOC is well documented on the band structures of these perovskites~\cite{even2013importance}. Intriguingly, SOC in conjunction with broken inversion symmetry acts as a key ingredient for several exotic phenomena such as persistent spin textures~\cite{tao2018persistent,zhao2020purely}, topological surface states~\cite{schindler2018higher} and Rashba-Dresselhaus (RD) effects~\cite{sandoval2012rashba,ponet2018first,di2016intertwined,PhysRevLett.115.037602,PhysRevB.95.245141}. 
 
 In absence of the inversion symmetry, the crystal feels an effective magnetic field due to SOC. This field coupled with spin moment leads to the momentum dependent splitting of bands, known as Rashba-Dresselhaus (RD) splitting. Dresselhaus \cite{dresselhaus1955spin} and Rashba \cite{rashba1960properties} effects were originally reported in zinc-blende and  wurtzite structures, respectively. The primary difference between both these effects resides in the origin of the noncentrosymmetry. In Rashba effect, there is site inversion asymmetry, whereas bulk inversion asymmetry is present in Dresselhaus effect. During the past decade, RD effect has been the subject of intense research due to its potential applications in the emergent field of spintronics~\cite{di2013electric,plekhanov2014engineering}. In view of this, lead halide perovskites have been studied as promising functional materials for spin based applications \cite{kim2014switchable,kutes2014direct,giovanni2015highly,kepenekian2017rashba,rinaldi2018ferroelectric}. Recently, a significant Rashba effect has been reported in tetragonal phase of MAPbI$_3$ (MA = CH$_3$NH$_3$${^+}$) due to the rotations of MA ion \cite{doi:10.1021/acs.jpclett.6b01794, leppert2016electric}. The effect has also been predicted in CsPbBr$_3$ and MAPbBr$_3$ perovskites \cite{isarov2017rashba,niesner2016giant}. Lately, there are several reports on ferroelectric-coupled Rashba effect in halide perovskites (viz. $\beta$-MAPbI$_3$, $\beta$-MASnI$_3$, ortho-MASnBr$_3$, FASnI$_3$ (FA = HC(NH$_2$)$_2$$^+$), opening new possibilities for perovskite based spin devices \cite{kim2014switchable, stroppa2014tunable}. However, due to the volatile nature of organic molecules, these perovskites suffer from poor stability towards heat and moisture~\cite{ouedraogo2020stability}. 
 
 Therefore, in our present work, we intend to explore inorganic FE perovskites in terms of RD splitting. Note that, the interplay between ferroelectricity and Rashba effect give rise to electric control of bulk Rashba effect, leading to an exciting class of Ferroelectric Rashba Semiconductors (FERSCs) \cite{picozzi2014ferroelectric,liebmann2016giant,PhysRevB.92.220101}. This effect can be theoretically quantified by employing $\textbf{\textit{k.p}}$ perturbation theory. According to this theory, the lowest-order RD Hamiltonian is given by \cite{manchon2015new, myung2018rashba}
 \begin{equation}
 	H_{{RD}} (\textbf{\textit{k}})=\alpha_{{R}}(\sigma_x k_y-\sigma_y k_x) + \alpha_{{D}}(\sigma_x k_x-\sigma_y k_y)
 \end{equation}
 Here, $\sigma_x$ and $\sigma_y$ are Pauli spin matrices, $k_i$ is the crystal momentum ($i= x, y, z$), $\alpha_{{R}}$ and $\alpha_{{D}}$ represent the Rashba and Dresselhaus coefficients, respectively. On solving the free-electron Hamiltonian with these terms, we get two spin-split polarized states having opposite spin polarization. Though Rashba and Dresselhaus exhibit a similar type of splitting, projection of spin in momentum space i.e., spin texture gives the idea on the nature of the splitting.
 Striving towards sizable Rashba-type splitting, CsPbF$_3$ perovskite may be a desirable candidate.  To the best of our knowledge, the available literature does not encompass the detailed quantitative study of RD effect in this material. 
 The presence of Pb and ferroelectricity in noncentrosymmetric phase of CsPbF$_3$ indicates the possibility of RD effect in this material. 
 
 Motivated by this idea, we have undertaken a theoretical study based on the perturbative $\textbf{\textit{k.p}}$ formalism and backed by the first-principles calculations. In the present work, we have studied \textit{Pm}$\bar{3}$\textit{m} and \textit{R}3\textit{c} phases of CsPbF$_3$. Firstly, we have examined the electronic band structures in the above mentioned phases. After that, we have estimated the band gap of these phases using first-principles based approaches combined with SOC, viz. density functional theory (DFT) \cite{hohenberg1964inhomogeneous,kohn1965self} with semi-local exchange-correlation ($\epsilon_{xc}$) functional (PBE \cite{perdew1996generalized}), hybrid DFT with HSE06 \cite{heyd2003hybrid, krukau2006influence} and single-shot GW (G$_0$W$_0$) \cite{hedin1965new, hybertsen1985first} under the many body perturbation theory (MBPT) and Wannier model approach. Subsequently, we have analyzed the electronic band structure of \textit{R}3\textit{c} phase in terms of Rashba splitting under the combined framework of DFT and perturbative $\textbf{\textit{k.p}}$ formalism. We have also investigated the effect of strain on electronic band gap and Rashba parameters of \textit{R}3\textit{c} phase. Finally, we have determined the minimum energy pathway of the FE transition using climbing image nudged elastic band (CINEB) method.
\begin{figure}[h]
	\includegraphics[width=0.95\textwidth]{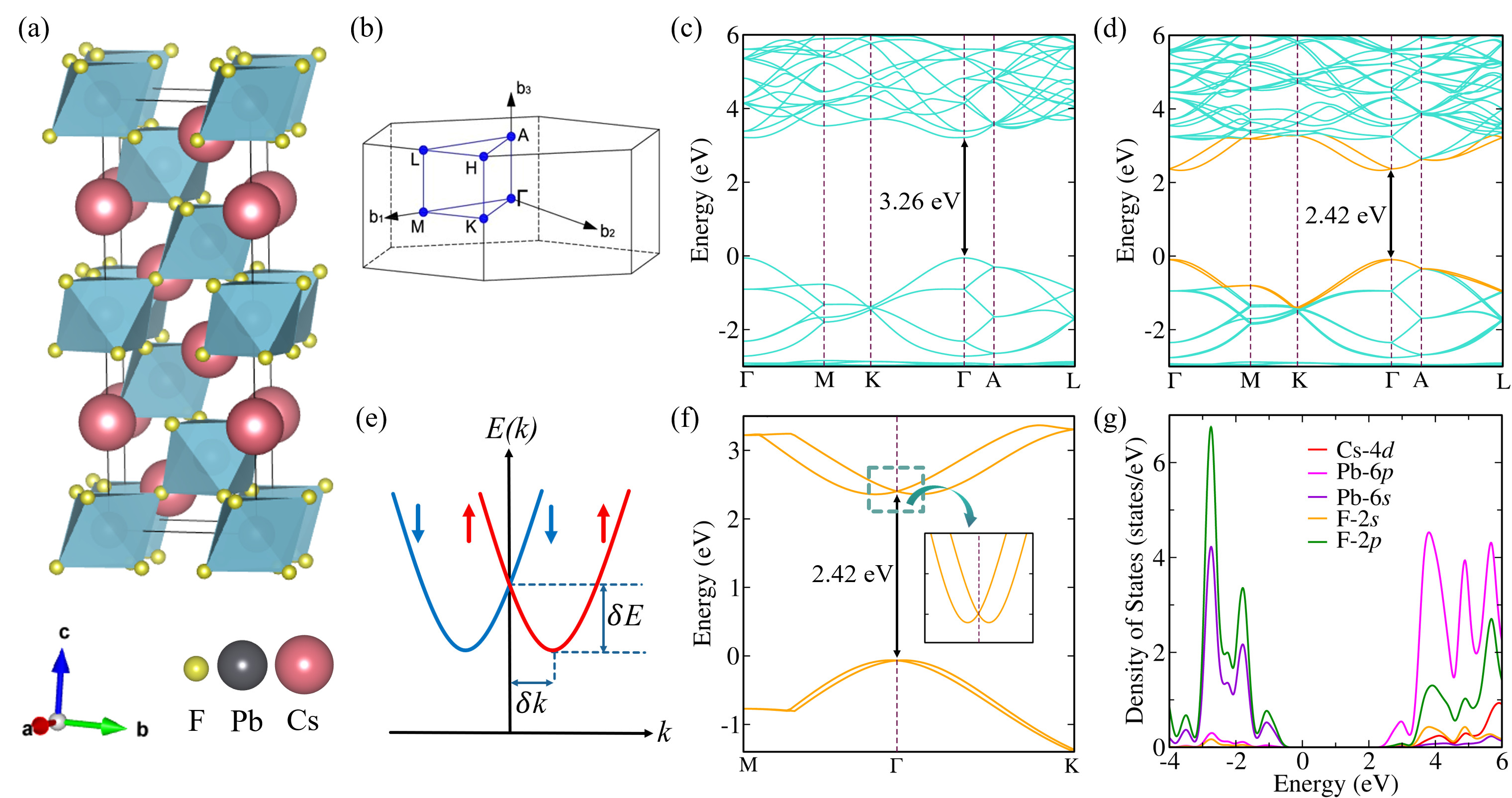}
	\caption{(a) Optimized crystal structure of CsPbF$_3$ in rhombohedral \textit{R}3\textit{c} phase. Cs, Pb and F atoms are indicated by red, grey and yellow colors, respectively. (b) The first hexagonal Brillouin zone showing the high symmetry path for band structure calculations in \textit{R}3\textit{c} phase of CsPbF$_3$. Electronic band structure of CsPbF$_3$ for \textit{R}3\textit{c} phase, calculated using (c) PBE (d) PBE+SOC. The conduction and valence bands considered in the discussion are indicated by orange color. (e) Schematic representation of bands showing Rashba splitting. (f) Splitting of conduction band minimum (CBm) and valence band maximum (VBM) of the chosen bands along the M-$\Gamma$-K path. The inset shows the enlarged view of CBm. (g) Projected density of states (pDOS) in \textit{R}3\textit{c} phase of CsPbF$_3$ calculated using HSE06+SOC. The Fermi energy is set to zero in the energy axis.}
	\label{pic2}
\end{figure}

CsPbF$_3$ mainly manifests in cubic (\textit{Pm}$\bar{3}$\textit{m}) and rhombohedral (\textit{R}3\textit{c}) phases~\cite{berastegui2001low, smith2015interplay}. The \textit{Pm}$\bar{3}$\textit{m} phase is centrosymmetric i.e., contains an inversion center. On the other hand, the rhombohedral \textit{R}3\textit{c} phase (see Fig. \ref{pic2}(a)) is noncentrosymmetric and exhibits FE behavior due to distortion of cations away from anionic polyhedra~\cite{smith2015interplay}. 
The calculated change in FE polarization of the rhombohedral (\textit{R}3\textit{c}) phase relative to the centrosymmetric structure is 34 $\mu$C/cm$^2$ along the [0001] direction in hexagonal setting (along the [111] direction in rhombohedral setting). The details of optimized crystal structures are provided in Sec. I of Supplemental Information (SI). 
Next, we have investigated the electronic band structures of  \textit{Pm}$\bar{3}$\textit{m} and \textit{R}3\textit{c} phases. The band structure of \textit{Pm}$\bar{3}$\textit{m} phase in presence of SOC reveals that there is no momentum-dependent splitting owing to its centrosymmetric structure (band structure and pDOS of \textit{Pm}$\bar{3}$\textit{m} phase are given in Sec. II of SI). In addition, the cubic \textit{Pm}$\bar{3}$\textit{m} phase is not dynamically stable (see Sec. III for phonon band structures). Therefore, we have explored RD splitting in the latter phase (i.e. \textit{R}3\textit{c}). We have plotted the band structure of FE \textit{R}3\textit{c} phase along the high symmetry path using hexagonal setting (as shown in Fig. \ref{pic2}(b)). Firstly, we have performed non-spin polarized calculations. After that we have considered SOC in the calculation of electronic band structure. Fig. \ref{pic2}(c) and \ref{pic2}(d) show the electronic band structures calculated using PBE and PBE+SOC, respectively. A direct band gap of 3.26 eV is observed without SOC, whereas on including SOC, the band gap is reduced to 2.42 eV (indirect) around the $\Gamma$ point. 
 This change in band gap is attributed to SOC arising from the presence of Pb-6$p$ orbitals in the conduction band (see pDOS in Fig. \ref{pic2}(g)). Hence, SOC is indispensable and duly considered in all further calculations. Note that PBE $\epsilon_{xc}$ functional is well known to underestimate the band gap due to its inability to capture the electron's self-interaction error. Therefore, we have also calculated the band gap with hybrid $\epsilon_{xc}$ functional HSE06 (with SOC), which comes out to be 3.57 eV. 
Until date, there is no theoretical or experimental report on the band gap of \textit{R}3\textit{c} phase. In view of this, for the better estimation of the band gap, we have employed G$_0$W$_0$ approximation on top of HSE06+SOC orbitals, that results in the band gap of 5.01 eV (see Fig. \ref{pic} (a)).
\begin{figure}[h]
	\centering
	\includegraphics[width=0.75\textwidth]{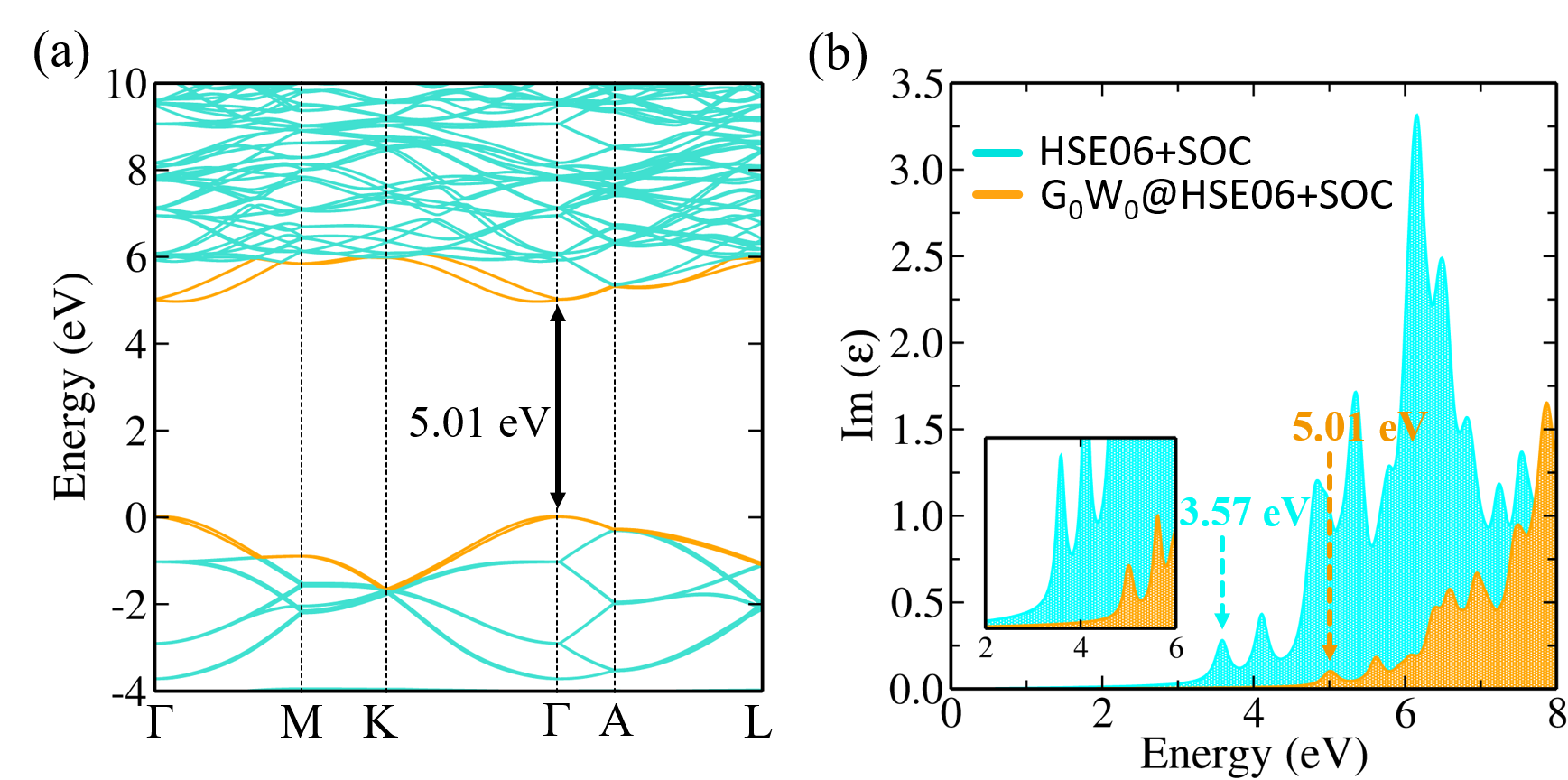}
	\caption{(a) Band structure of $R3c$ phase using G$_0$W$_0$@HSE06+SOC. (b) Imaginary part (Im ($\varepsilon$)) of the dielectric function for \textit{R}3\textit{c} phase of CsPbF$_3$ calculated using HSE06+SOC and G$_0$W$_0$@HSE06+SOC.}
	\label{pic}
\end{figure}
\begin{table}[htbp]
	\caption {Band gap (in eV) of $Pm\bar{3}m$ and $R3c$ phases using different $\epsilon_{xc}$ functionals.}
	\begin{center}
		\begin{adjustbox}{width=0.8\textwidth} 
			\setlength\extrarowheight{+4pt}
			\begin{tabular}[c]{|c|c|c|c|c|c|} \hline		
				\textbf{Structure} & \textbf{PBE} & \textbf{PBE+SOC} & \textbf{HSE06} & \textbf{HSE06+SOC} & \textbf{G$_0$W$_0$@HSE06+SOC} \\ \hline
				$Pm\bar{3}m$        &  2.62  & 1.69 & 3.61\cite{qian2016comprehensive, murtaza2011structural} & 2.71     &        4.38      \\ \hline
				$R3c$        &  3.26  &  2.42  & 4.34 & 3.57     &        5.01        \\ \hline
			\end{tabular}
		\end{adjustbox}
		\label{T2}
	\end{center}
\end{table}
Subsequently, we have also plotted the imaginary part (Im ($\varepsilon$)) of dielectric function with converged \textit{k}-grid (for \textit{k}-grid convergence, see Sec. IV in SI) for \textit{R}3\textit{c} phase of CsPbF$_3$ using HSE06+SOC and G$_0$W$_0$@HSE06+SOC (see Fig. \ref{pic} (b)). The first peak of Im ($\varepsilon$) corresponds to the electronic band gap, which is the same as we get from band structure calculations. For imaginary part (Im ($\varepsilon$)) of dielectric function of \textit{Pm}$\bar{3}$\textit{m} phase, see Sec. V in SI. The values of band gaps are listed in Table \ref{T2}. It is worth quoting that HSE06+SOC/G$_0$W$_0$@HSE06+SOC only enhances the band gap without any notable change in the nature of band structure and strength of Rashba splitting \cite{di2013electric, doi:10.1021/acs.jpclett.6b01794}. To confirm this, we have compared the band structures of \textit{Pm}$\bar{3}$\textit{m} phase using PBE+SOC, HSE06+SOC, G$_0$W$_0$@HSE06+SOC and have found that the band profile remains same (see Sec. VI in SI for details).  
Therefore, in view of computational cost PBE+SOC is considered to compute RD parameters.\\
\indent Inclusion of SOC leads to the splitting of bands in $k_z$=$\frac{\pi}{c}$ plane, which is perpendicular to the direction of polarization (see Fig. \ref{pic2}(d)). On the contrary, splitting is completely absent along $\Gamma$-A i.e., in direction parallel to polarization axis, indicating that the momentum dependent splitting around $\Gamma$ point is the Rashba-type splitting. As a consequence, the CBm and VBM are located slightly off the $\Gamma$ point. The CBm shifts from $\Gamma$ towards M and K by 0.075 \AA$^{-1}$ and 0.061 \AA$^{-1}$, respectively. The VBM shifts from $\Gamma$ towards M and K by 0.022 \AA$^{-1}$ and 0.012 \AA$^{-1}$, respectively 
(see Fig. \ref{pic2}(f)). This shift from $\Gamma$ point in either direction is known as offset momentum ($\delta$$k$) (see Fig. \ref{pic2}(e)). The difference of energies at $\Gamma$ and extremum point is known as Rashba spin-splitting energy ($\delta$$E$).

\begin{figure*}[h]
	\includegraphics[width=0.95\textwidth]{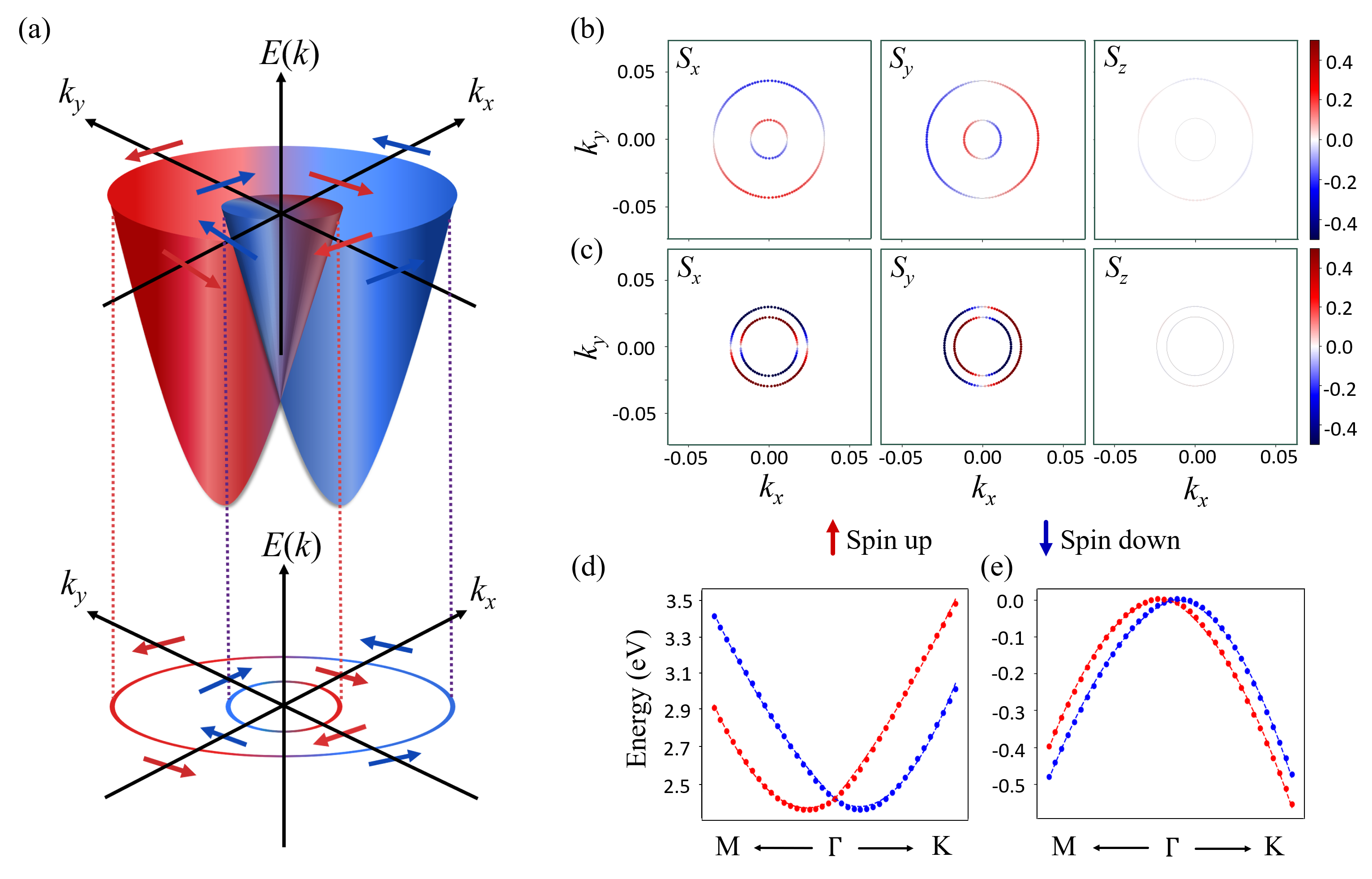}
	\caption{(a) Schematic representation of spin textures in Rashba splitting. Spin projected constant energy contour plots of spin texture calculated in $k_x$-$k_y$ plane centered at $\Gamma$ point. The upper and lower panels represent the spin textures calculated at constant energy: (b) E=E$_\textrm{F}+2.5$ eV and (c) E=E$_\textrm{F}-0.2$ eV, respectively. Electronic band structures showing spin-splitting of (d) CBm and (e) VBM around $\Gamma$ point, respectively. Band structure is plotted along ($\frac{2\pi}{a}0.25, 0, 0$) $-$ (0, 0, 0) $-$ ($\frac{2\pi}{a}0.16, \frac{2\pi}{b}0.16, 0$) direction of momentum space, which is M $-$ $\Gamma$ $-$ K direction. DFT and $\textbf{\textit{k.p}}$ band structures are plotted with dashed lines and dots, respectively. In the color scale, red depicts spin-up while blue depicts spin-down states.}
	\label{pic4}
\end{figure*}

To grasp the overall nature of splitting, the constant energy 2D contour plots of spin texture are plotted in $k_x$-$k_y$ plane centered at $\Gamma$ point (see Fig. \ref{pic4}(a) for schematic representation of spin textures).
Fig. \ref{pic4}(b) and \ref{pic4}(c) show $x$, $y$ and $z$ components of spin at constant energies around CBm and VBM, respectively. As we can see from spin textures of CBm and VBM, the in-plane spin components (S$_x$ and S$_y$) are perpendicular to the crystal momentum and out-of-plane component (S$_z$) is completely absent. This results in the helical shape of spin texture with inner and outer bands having opposite orientation, which confirms the existence of dominant Rashba splitting. 
In order to have a quantitative study of the RD effect, we have considered the $\textbf{\textit{k.p}}$ Hamiltonian. The model Hamiltonian of $R3c$ structure possessing $C_{3v}$ point group symmetry near $\Gamma$ point in the presence of SOC can be written as \cite{PhysRevB.101.014109, PhysRevB.97.205138, PhysRevB.93.245159}:
\begin{equation}
\begin{aligned}
\begin{split}
H_{\Gamma}(\textbf{\textit{k}}) & =  H_o(\textbf{\textit{k}})+\alpha \sigma_yk_x+\beta \sigma_xk_y + \gamma \sigma_z [(k^3_x+k^3_y)-3(k_xk^2_y+k_yk^2_x)]
\label{eq3}
\end{split}
\end{aligned}
\end{equation} 
Here $H_o(\textbf{\textit{k}})$ is the Hamiltonian of the free electrons with eigenvalues $E_o(\textbf{\textit{k}})$. $\alpha$, $\beta$ are the coefficients of the linear term and $\gamma$ is the coefficient of the higher order term in the Hamiltonian. Since we have not noticed out-of-plane spin component in the spin texture, therefore, we have neglected the higher order terms in the Hamiltonian (see Sec. VII in SI for more details). Hence, the Hamiltonian for $\Gamma$ point (considering only linear terms) can now be written as:
\begin{equation}
	\begin{aligned}
		\begin{split}
			H_{\Gamma} (k) & =  H_o(k)+\alpha k_x\sigma_y+\beta k_y\sigma_x \\& = H_o(k)+\alpha_R (\sigma_x k_y-\sigma_y k_x)\\& + \: \alpha_D (\sigma_x k_x-\sigma_y k_y)
			\label{eq3}
		\end{split}
	\end{aligned}
\end{equation}

\noindent The energy eigen values corresponding to this Hamiltonian is given as, $E_o(k)=\frac{\hbar^2k_x^2}{2m^*_x}+\frac{\hbar^2k_y^2}{2m^*_y}$ and $m^*_x$, $m^*_y$ are the effective masses in $x$ and $y$ directions, respectively. Rashba and Dresselhaus coefficients are defined as $\alpha_R= (\alpha-\beta)/2$ and $\alpha_D= (\alpha+\beta)/2$, respectively \cite{sheoran2021rashba}. On diagonalizing Eq. \ref{eq3}, we get energy eigenvalues as:
\begin{equation}
	E_{\pm}({k})= E_o(k) \pm \sqrt{\alpha^2 k_x^2+\beta^2 k_y^2}
\end{equation}
The band structures obtained from DFT and the model Hamiltonian around $\Gamma$ point are shown with dashed lines and dots in Fig. \ref{pic4}(d) and \ref{pic4}(e), respectively. The good agreement between both the band structures justifies the reliability of the chosen Hamiltonian. The DFT calculations for CBm give $\delta E$ = 43.5 meV, ${\delta k}$ = 0.075 \AA$^{-1}$ along $\Gamma$-M direction, where $k_y$ = 0. Hence, the value of $\alpha$ = $\frac{2 \delta E}{\delta k}$ = 1.16 eV$\textrm{\AA}$. Along $\Gamma$-K direction, where $k_x$ = $k_y$, $\delta E$ = 43.5 meV, ${\delta k}$ = 0.061 \AA$^{-1}$ and the calculated value of $\sqrt{\alpha^2+\beta^2}$ = $\frac{2 \delta E}{\delta k}$ = 1.43 eV$\textrm{\AA}$. On putting the value of $\alpha$, we get $\beta$ = --0.84 eV$\textrm{\AA}$, which results in $\alpha_R$ = 1.00 eV$\textrm{\AA}$ and $\alpha_D$ = 0.16 eV$\textrm{\AA}$. Similarly, for VBM, $\delta E$ = 3.4 meV, ${\delta k}$ = 0.022 \AA$^{-1}$ along $\Gamma$-M direction. So, the value of $\alpha$ = $\frac{2 \delta E}{\delta k}$ = 0.31 eV$\textrm{\AA}$. Along $\Gamma$-K direction, $\delta E$ = 3.4 meV and ${\delta k}$ = 0.012 \AA$^{-1}$. Therefore, $\sqrt{\alpha^2+\beta^2}$ = $\frac{2 \delta E}{\delta k}$ = 0.57 eV$\textrm{\AA}$, which gives $\beta$ = --0.48 eV$\textrm{\AA}$. Using $\alpha$ and $\beta$ parameters, the estimated values of $\alpha_R$ and $\alpha_D$ are 0.40 eV$\textrm{\AA}$ and $\alpha_D$ = 0.09 eV$\textrm{\AA}$, respectively. The values are summed up in Table \ref{T3}. As we can see from the values of RD parameters, the Rashba effect dominates in the conduction band. The origin of large Rashba effect in CBm can be attributed to the stronger SOC stemming from higher contribution of Pb-6$p$ orbital in conduction band (see Fig. \ref{pic2}(g)). This large contribution of SOC in CBm compared to VBM is in well agreement with the previous findings \cite{stranks2018influence}. On fitting the DFT band structure around $\Gamma$ point for CBm, we find $\alpha$ = --0.90 eV$\textrm{\AA}$ and $\beta$ = 1.20 eV$\textrm{\AA}$. This gives $\alpha_R$ = 1.05 eV$\textrm{\AA}$ and $\alpha_D$ = 0.15 eV$\textrm{\AA}$. Similarly, fitting the band structure for VBM, gives $\alpha$ = --0.33 eV$\textrm{\AA}$ and $\beta$ = 0.49 eV$\textrm{\AA}$ and hence $\alpha_R$ = 0.41 eV$\textrm{\AA}$ and $\alpha_D$ = 0.08 eV$\textrm{\AA}$. These values are in well agreement with the predicted values based on DFT calculation. The calculated values of RD parameters also confirm that the splitting around $\Gamma$ point is mainly dominated by the Rashba effect.
\begin{figure*}[h]
	\includegraphics[width=1\textwidth]{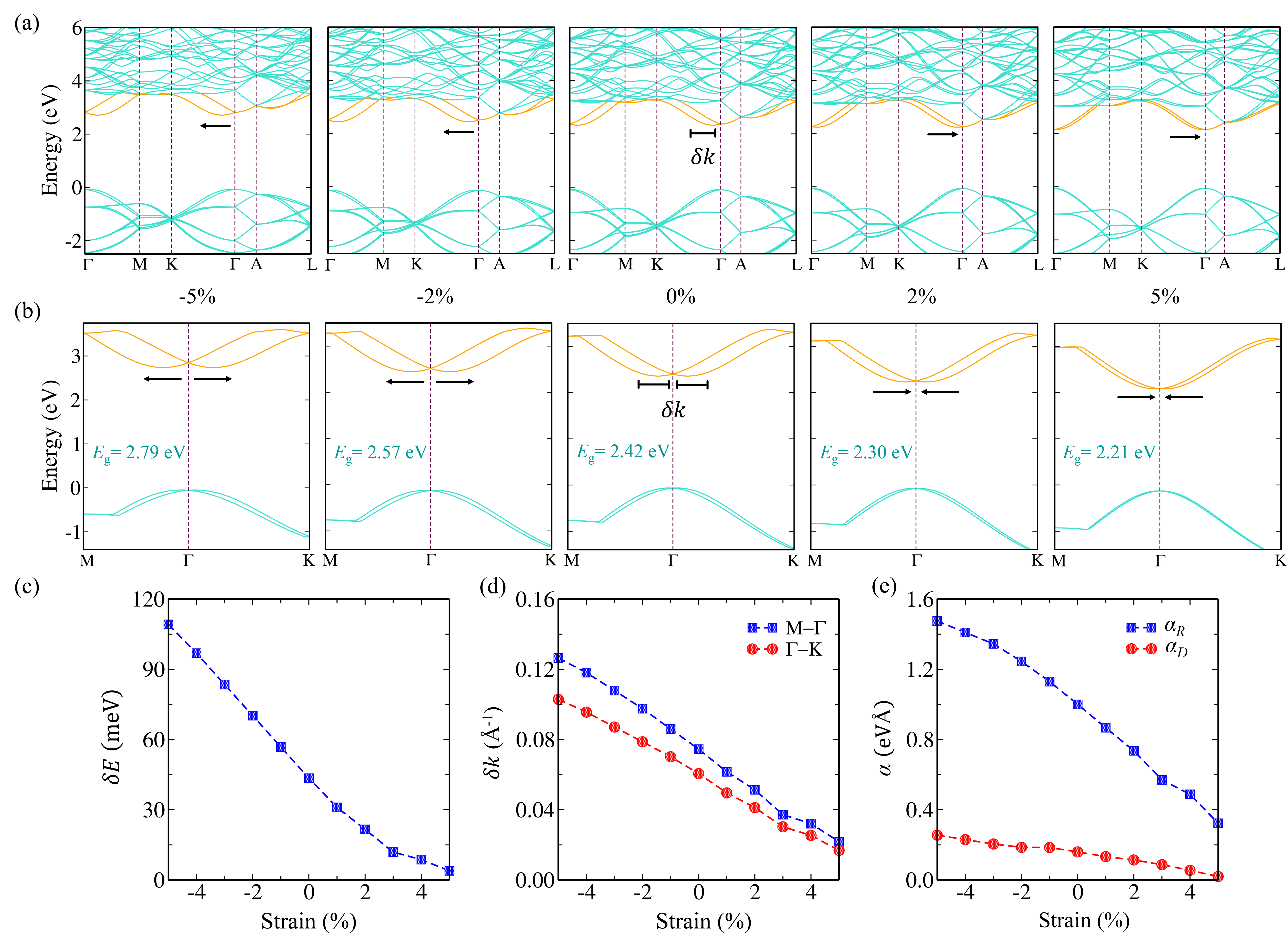}
	\centering
	\caption{Band structures of CsPbF$_3$ ($R3c$ phase)  under uniaxial strain of $\pm$5\%, calculated using PBE+SOC for path (a) $\Gamma$-M-K-$\Gamma$-A-L and (b) M-$\Gamma$-K. (c) Rashba spin-splitting energy ($\delta E$) (d) offset momentum ($\delta k$) (e) Rashba parameters ($\alpha_R$ and $\alpha_D$) as a function of strain. The values are calculated for CBm in $R3c$ phase. Note that $\delta E$ is same along both M-$\Gamma$ and $\Gamma$-K directions.}
	\label{pic5}
\end{figure*}

\begin{table}[htbp]
	\caption {Rashba parameters for band-splitting at $\Gamma$ point in $R3c$ phase.}
	\begin{center}
		\begin{adjustbox}{width=0.8\textwidth} 
			\setlength\extrarowheight{+4pt}
			\begin{tabular}[c]{|c|c|c|c|c|c|} \hline		
				\textbf{Position} & \textbf{$\delta E$ (meV)} & \textbf{$\delta$\textit{k}$_{\Gamma-\textrm{M}}$ (\AA$^{-1}$)} & \textbf{$\delta$\textit{k}$_{\Gamma-\textrm{K}}$ (\AA$^{-1}$)} & $\alpha_R$ \textbf{(eV\AA)} & \bf $\alpha_D$ (eV\AA) \\ \hline
				CBm        &  43.5  & 0.075 & 0.061& 1.05     &        0.15       \\ \hline
				VBM        &  3.4  &  0.022  & 0.012 & 0.41     &        0.08        \\ \hline
			\end{tabular}
		\end{adjustbox}
		\label{T3}
	\end{center}
\end{table}

After thorough analysis of RD spin-splitting, we have investigated the effect of strain on the band structure of $R3c$ phase. For this, we have applied uniaxial strain in $z$-direction, which is defined as
\begin{equation}
	x=\frac{c-c_0}{c_0} \times 100\%
\end{equation}  
where $c_0$ is the equilibrium lattice constant and $c$ is the strained lattice constant. The lattice vector $c$ is varied from -5\% to +5\%, where ``+'' and ``$-$'' are used to denote tensile and compressive strain, respectively. After optimizing the structures at a given strain, we have plotted the band structures along high symmetry path. Fig. \ref{pic5}(a) shows the band structure of $R3c$ phase along $\Gamma$-M-K-$\Gamma$-A-L path, under the uniaxial strain of $\pm$5\%. To clearly examine the shift, we have also plotted the band structures along M-$\Gamma$-K (see Fig. \ref{pic5} (b)). Here, we have focused on CBm as there is large Rashba splitting in comparison to VBM. From the band structures in Fig. \ref{pic5} (b), we infer that strain causes a shift in momentum ($\delta k$) on either side of $\Gamma$ point. The arrows represent the direction of shift. We have taken the equilibrium band structure (i.e., with 0\% strain) as the reference for all the strained band structures. Interestingly, under compressive strain, the bands shift off from $\Gamma$ on either side towards M and K. On the contrary, under tensile strain, the bands shift towards $\Gamma$ from either side. This in turn changes $\delta k$ and $\delta E$ as a function of strain, and the overall effect of these parameters will change $\alpha_R$ and $\alpha_D$ accordingly. To quantify the effect of strain on Rashba parameters, we have calculated their values at a given strain within the framework of DFT. Fig. \ref{pic5} (c-e) show $\delta E$, $\delta k$, $\alpha_R$ and $\alpha_D$ as a function of strain (for details, see Sec. VIII in SI). From Fig. \ref{pic5} (c), we have observed that under compressive strain $\alpha_R$ is significantly enhanced from 1.05 to 1.48 eV{\AA} making the material tunable for spintronics application.
Also, we have seen notable change in electronic band gap on the application of strain. The band gap values are mentioned in Fig.\ref{pic5}(b) (for more quantitative details, see Sec. IX in SI). 
\begin{figure}[H]
	\centering
	\includegraphics[width=0.7\textwidth]{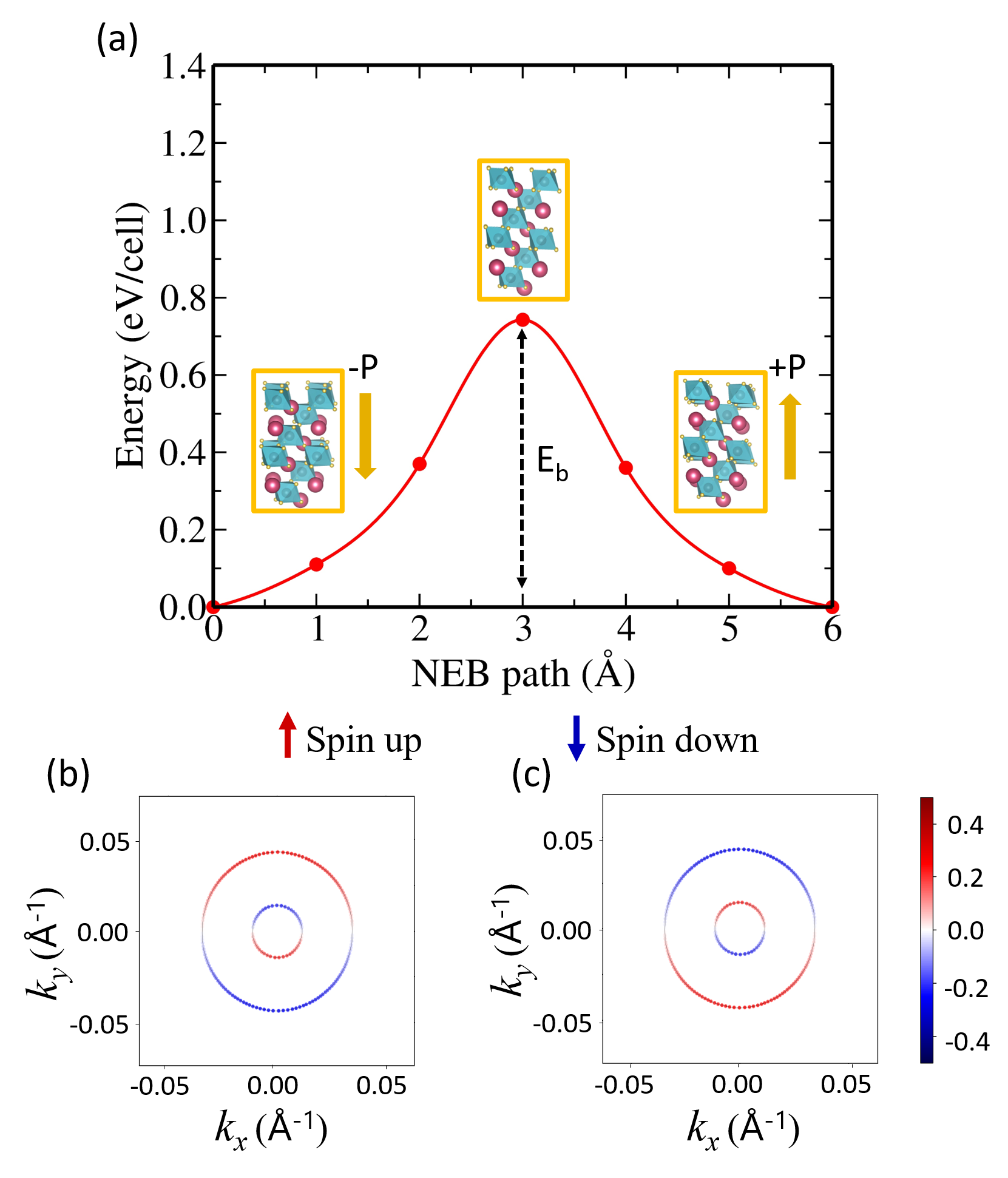}
	\caption{(a) Climbing image nudged elastic band (CINEB) calculation for the polarization switching process in CsPbF$_3$ perovskite. Two ferroelectric (FE) structures in the ground state with opposite direction of electric polarization are shown. E$_b$ is the activation barrier energy for the polarization switching process. Reversible in-plane spin textures calculated at constant energy E=E$_F$+2.5 eV with oppposite spin polarization: (b) -P (c) +P.}
	\label{pic}
\end{figure}
Lastly, we have explored the possibility of polarization switching in FE CsPbF$_3$ using strain effects. To predict the feasibility of this phenomenon, we have analyzed the minimum energy pathway of the FE transition using CINEB method \cite{PhysRevB.100.115104, PhysRevLett.122.116401}. For this, we have chosen the structure deformation path between two FE states with opposite spontaneous polarization through a centrosymmetric reference. As shown in Fig.\ref{pic}, the E$_b$ denotes the activation barrier for the polarization switching, which comes out to be 0.75 eV. This suggests that a switchable FE polarization is plausible in the material. The strain tunability and reversible spin textures owing to FE switching in CsPbF$_3$ give electrical control of spins and provide an all-semiconductor design for spintronic devices such as in spin-field effect transistor \cite{kepenekian2015rashba, chen2021tunable}.




In summary, using first-principles based methodologies viz. PBE, HSE06 and many-body perturbation theory (G$_0$W$_0$), we have systematically studied the electronic structure of \textit{Pm}$\bar{3}$\textit{m} and \textit{R}3\textit{c} phases of CsPbF$_3$. Note that SOC is duly considered in all the calculations. The \textit{R}3\textit{c} phase being noncentrosymmetric is investigated in terms of RD splitting. The DFT calculations are well corroborated by symmetry adapted $\textbf{\textit{k.p}}$ model Hamiltonian. The helical nature of in-plane spin components in spin textures indicates the Rashba-type splitting. The calculated values of RD parameters suggest that the Rashba effect is predominant in our material while Dresselhaus effect has a negligible contribution. In addition, a larger Rashba effect is observed in CBm than VBM of \textit{R}3\textit{c} phase owing to the larger contribution of Pb-6$p$ orbital in conduction band. Further, 
we have noticed that the Rashba parameters viz. $\delta k$, $\delta E$, $\alpha_R$, and $\alpha_D$ almost linearly increase under compressive strain. The band gap of CsPbF$_3$ is significantly tuned with the application of strain. Also, we have observed reversible spin textures via FE switching. These results indicate that FE CsPbF$_3$ is having enough potential as an excellent amenable material in the field of spintronics. 

\section{Computational Methods}
The DFT based first-principles calculations have been performed using the Vienna \textit{ab initio} simulation package (VASP) \cite{kresse1993ab, kresse1996efficient}. The ion-electron interactions in all the elemental constituents are described using projector augmented wave (PAW)  \cite{blochl1994projector, kresse1999ultrasoft} method as implemented in VASP. We have considered pseudopotentials with the following valence states: Cs, 5s$^2$5p$^6$6s$^1$; Pb, 6s$^2$5d$^{10}$6p$^2$; F, 2s$^2$2p$^5$. The structures are optimized using Perdew-Burke-Ernzerhof generalized gradient approximation (PBE-GGA), relaxing all ions until Hellmann-Feynman forces are less than 0.001 eV/\AA. The cutoff energy of 520 eV is used for the plane-wave basis set such that the total energy calculations are converged within 10$^{-5}$ eV. The $\Gamma$-centered 6$\times$6$\times$6 and 9$\times$9$\times$4 \textit{k}-grid are used to sample the irreducible Brillouin zones of cubic phase with \textit{Pm}$\bar{3}$\textit{m} space group and rhombohedral phase with \textit{R}3\textit{c} space group of bulk CsPbF$_3$, respectively. The phonon calculations are performed for 3$\times$3$\times$3 and 2$\times$2$\times$2 supercells in \textit{Pm}$\bar{3}$\textit{m} and \textit{R}3\textit{c} phases using PHONOPY package \cite{togo2008first, togo2015first}. In order to predict band gap, single-shot GW (G$_0$W$_0$) calculations have been performed on top of the orbitals obtained from HSE06+SOC $\epsilon_{xc}$  functional [G$_0$W$_0$@HSE06+SOC]. For this, we have used 6$\times$6$\times$6 and 4$\times$4$\times$2 \textit{k}-grids in \textit{Pm}$\bar{3}$\textit{m} and \textit{R}${3}$\textit{c} phases, respectively. The number of bands is set to four times the number of occupied bands.  The polarizability calculations are performed on a grid of 50 frequency points. FE polarization is evaluated in the framework of Berry-phase theory of polarization \cite{king1993theory, resta1994macroscopic}. The minimum energy pathways of FE transitions are determined through the climbing image nudged elastic band (CINEB) method \cite{henkelman2000climbing}. Spin textures are plotted by calculating expectation values of spin operators $S_{i}$ (\textit{i} = \textit{x}, \textit{y}, \textit{z}), given by
\begin{equation}
	\langle S_{i} \rangle = \frac{1}{2} \bra { \Psi_k} \sigma_i \ket {\Psi_k}
\end{equation}
where $\sigma_i$ are the pauli matrices and $\Psi_k$ is the spinor eigenfunction obtained from noncollinear spin calculations. The spin texture is calculated with closely spaced 12$\times$12 \textit{k}-grid around high symmetry points. The PyProcar code is used to calculate the constant energy contour plots of the spin texture \cite{HERATH2020107080}. \\
\begin{acknowledgement}
P.B. acknowledges UGC, India, for the senior research fellowship [grant no. 1392/(CSIR-UGC NET JUNE 2018)]. D.G. acknowledges UGC, India, for the senior research fellowship [grant no. 1268/(CSIR-UGC NET JUNE 2018)]. S.S. acknowledges CSIR, India, for the junior research fellowship [grant no. 09/086(1432)/2019-EMR-I]. S.B. acknowledges the financial support from SERB under core research grant (grant no. CRG/2019/000647). We acknowledge the High Performance Computing (HPC) facility at IIT Delhi for computational resources.
\end{acknowledgement}
\begin{suppinfo}
Optimized structures of \textit{Pm}$\bar{3}$\textit{m} and \textit{R}${3}$\textit{c} phases; Band structure and pDOS (projected density of states) of cubic \textit{Pm}$\bar{3}$\textit{m} phase; \textit{k}-grid convergence in \textit{Pm}$\bar{3}$\textit{m} and \textit{R}${3}$\textit{c} phases; Imaginary part (Im($\varepsilon$)) of the dielectric function for cubic \textit{Pm}$\bar{3}$\textit{m}  phase; Comparison of PBE+SOC and HSE06+SOC band structures of cubic \textit{Pm}$\bar{3}$\textit{m} phase; RD parameters as a function of strain in \textit{R}${3}$\textit{c} phase; Band gap as a function of strain in \textit{R}${3}$\textit{c} phase.
	
\end{suppinfo}
\bibliography{achemsodemo.bib}